# A Conjecture About Fermi-Bose Equivalence

Paul J. Werbos, pwerbos@nsf.gov[1]

The purpose of this note is to ask about the possibility of an exact equivalence between: (1) a bosonic quantum field theory (BQFT) which yields topological solitons, similar to the classical Skyrme model ; and (2) a standard mixed fermi/bose QFT , such as QED or the standard model of physics -- *in the limit* where the radius r of the topological soliton goes to zero. I am not asserting that this does work, but there are reasons to hope that it *might*, and there are very good reasons to believe it would be very important if true.

## 1. Background

Classic papers by Coleman [1] and Mandelstam [2] proved the equivalence of the quantum sine-Gordon (QSG) model – a bosonic 1+1-D model – to the Massive Thirring Model (MTM), a purely fermionic model. The QSG model was essentially the 1+1-D member of Skyrme's family of models embodying the concept of *topological charge* [3]. Mandelstam proved the equivalence by first proposing a "definition" of the fermionic field operators of MTM as functions of the bosonic field operators of.QSG. He then proved that they have the right anticommutation relations and that substituting these definitions into the Hamiltonian of MTM gives us back the Hamiltonian of QSG. Coleman expressed some initial shock that it seems we cannot tell the difference between a fermion and a boson here; however, he explained this by noting that the elementary fermion of MTM appears as a bound state of the elementary bosons of QSG and vice-versa.

Wilczek and Zee [4] proved a similar equivalence in 2+1-D, which is perhaps the most important result so far in more than one spatial dimension. This equivalence has been studied in great depth, and applied very widely in solid state physics and electronics [5,6]. Xiao-Gang Wen (in section 6.4 of [6]) provides an excellent summary of the key result, as follows. Consider the bosonic field theory governed by the Lagrangian

$$\mathcal{L} = \frac{\chi}{2}(\partial_\mu \mathbf{n})^2, \qquad (1)$$

where $\mathbf{n}$ is a unit vector in $R^3$. If we define the two-spinor $z^T=(z_1, z_2)$ and choose:

$$\mathbf{n} = z^\dagger \boldsymbol{\sigma} z, \qquad z^\dagger z = 1 \qquad (2)$$

and define:

$$c_\mu = -(i/2)(z^\dagger \partial_\mu z - (\partial_\mu z)^\dagger z), \qquad (3)$$

we obtain an equivalent Lagrangian:

$$\mathcal{L} = 2\chi \left|(\partial_\mu - ic_\mu)z\right|^2 \qquad (4)$$

Equation 1 is a simple bosonic topological soliton model, in the family considered by Skyrme [3,7]. Equation 4 is a mixed fermi-bose system, where $c_\mu$ is an emergent bosonic gauge field.

The obvious way to generalize this to 3+1-D would be to replace equation 1 by a full 3+1-D model, such as based on a bosonic (isospinor) field $\varphi^\mu$ in $R^4$ and to represent $\varphi^\mu$ as:

$$\varphi^\mu = \psi_\alpha^\dagger \gamma^\mu_{\alpha\beta} \psi_\beta, \qquad (5)$$

where $\psi$ is a Dirac 4-spinor and $\gamma^\mu$ are the usual Dirac matrices. But research on bosonization has shown that the simple kind of exact equivalence found in 2+1-D does not apply here [8].

---

[1] The views herein are the personal views of the author, and do not reflect official views of anyone

## 2. The Conjecture

There is a simple intuitive explanation for *why* the exact equivalence breaks down in 3+1-D. Let us denote the set of parameters of the Skyrme model or one of its relatives as W. When W defines a nice, continuous field theory with soliton solutions, we expect these solutions to have a nonzero radius r. But the 3+1-D version of equation 4 would be equivalent to something more like the Maxwell-Dirac system, which assumes particles of radius zero. *Prior to renormalization*, the Maxwell-Dirac system yields particles (eigenvectors) of infinite mass-energy, because of the infinite energy of self-repulsion which results from concentrating a charge at a point. But [1,7] the quantum solitons of the system have mass-energy less than the finite masses of their classical counterparts, for any reasonable r>0.

The conjecture, then, is that standard Fermi-bose QFTs are equivalent to the *limit* of an interesting BQFT model, as r goes to zero, when the parameters of the BQFT model move along some path W(r). I will state this more precisely below. One obvious choice for the BQFT would be the system studied by 'tHooft and Polyakov, or the system defined by Hasenfratz and 'tHooft[9]. (These avoid the need to assume unit length for the spinor.) Perhaps the work of Wen [6], of Laughlin or of Penrose [10] may suggest alternative possibilities. If the equivalence ends up mapping three covariant vectors into a set of three Dirac spinors plus associated gauge fields, this may be a good start towards developing a BQFT which would map into the entire standard model of physics.

Glossing over that important issue, I would propose the following sort of equivalence between the BQFT model and a mixed Fermi-Bose QFT..

Let us define $H_\psi(r)$ as a *regularized and renormalized* version of the Hamiltonian of the mixed fermi-bose QFT being mapped into. Let r be a regularization parameter, such that the limit $H_\psi(0)$ corresponds exactly to the traditional QFT being mapped into. (In addition, to make this completely rigorous in the *spirit* of axiomatic quantum field theory[11], we may "smear" the creation and annihilation operators by some factor like r/10**6, to make sure that they are well-defined for all r>0.) Crudely, let us try to pick a regularization scheme such that $H_\psi(r)$ differs from $H_\psi(0)$ only on a distance scale less than r; this is just a heuristic rule to try to find *a* suitable regularization scheme, and we do not need to worry about characterizing all possible choices of regularization if we find at least one that works.

Let us define $H_\psi^{(\varphi)}(r)$ as the operator which results when we substitute equation 5 into $H_\varphi(r)$, the Hamiltonian of the BQFT model for parameters W(r), *without* regularization/renormalization. In effect, the hope is that varying a mass and coupling term as functions of r can take the place of regularization. Following the approach of Wilczek and Zee [4], we would want to add $\theta H_J$ to $H_J$ to $H_\psi^{(\varphi)}(r)$, where $H_J$ is a conserved quantity associated with the topological charge and $\theta$ is an arbitrary scalar which we can choose so as to make the equivalence easy to prove. As in Wilczek and Zee, the gauge fields in the Fermi-Bose QFT would be created, in effect, by this term.

In addition, let us define $\Omega^\chi$ as the set of possible wave functions (unit vectors in the usual Fock-Hilbert space of QFT) made up of the Fock space analog of ordinary functions whose size (norm), gradient and integrals of the same, and energy, are all bounded by the real number $\chi$. (The Fock space analog is to bound the operator expectation values, and here we can bound by both energy measures.) Strictly speaking, this strong definition of $\Omega^\chi$ would be very useful for initial analysis, but would need to be extended to handle scattering states [12] in a proper way; the axioms of Streater and Wightman have the same problem, insofar as the concept of Hilbert space requires $L^2$-boundedness, but they cite work by Ruelle and Haag as one way (at least) to make the kind of extension needed for true axiomatic rigor in the complete analysis of scattering probabilities.

Given these definitions, I can now state the conjecture.

The conjecture is that there exists a suitable path of parameters W(r) and of regularization and renormalization such that: For any wave function $\psi$ which is in $\Omega^\chi$ for some $\chi$,

$$\left|\left(H_\psi(r) - H_\psi^{(\varphi)}(r)\right)\psi\right| \to 0 \quad as \quad r \to 0 \qquad (6)$$

The reason to hope this is that the soliton goes to a point particle as r goes to zero, exactly like the electron. On any given frequency scale (roughly determined by $\chi$), the difference between the limit and the regularized versions becomes more and more "invisible" as it becomes concentrated into a smaller and smaller volume of space. The gauge field of the BQFT model becomes more and more like the usual gauge fields of the standard model.

The same spin effects which yield a fermionic type of solution in 2+1-D would also be at work here, for the asymptotic values of the field away from the point particle core [3].

As in Coleman's analysis, we may find an exact correspondence between bound states in one theory and elementary solitons in another. However, certain solitons of the BQF (like monopoles) may diverge to infinite mass, to particles with no empirical consequences and no disturbance to the equivalence of predictions. Others would converge to each other *as tempered distributions*, as functions which yield the same inner products with any possible ψ in $\Omega^\chi$ for some χ.

## 3. Potential Significance *If True*

*If* this conjecture (or a modified version of it) should be true, there would be a long chain of consequences of great potential importance. For example, consider the BQFT model for W(r) with r>0 less than the known empirical upper bound for the radius of the electron (currently 10**-18 cm, much larger than the Planck length radius of the electron assumed in superstring theory). This theory would be indistinguishable as yet from the mixed fermi-bose theory, *and yet* it would be a well-defined "finite" theory. More precisely, there is good reason to believe that this theory would "exist" in some axiomatic sense (albeit not in the precise sense of Streater and Wightman [11]) and would be well-defined even without renormalization! However, it would be necessary to use special methods used to analyze soliton-bearing theories (e.g. [7]) and related nonperturbative methods in order to prove this.

The axioms of Streater and Wightman call for the use of conventional $L^2$ norms in proving the existence (well-definedness) of a quantum field theory.. But the dynamics of traditional QFT does not behave so nicely with respect to those traditional norms. For example, the "problem" of "infrared catastrophe" in quantum electrodynamics is essentially a red herring, an example of good behavior which looks bad when one uses a norm which does not fit well to the natural dynamical flows of the system under study. Equation 6 and the concept of $\Omega^\chi$ implicitly provide a different kind of norm. Using this kind of norm, one may reasonably hope to prove that the BQFT is a well-defined dynamical system by exploiting the analogy to rigorous work on the existence of classical dynamical systems [13]. Because time is given a special role in this analysis, it would not immediately meet Streater and Wightman's goal of avoiding such a special role; however, the next to last paragraph below will suggest a way to meet that goal, in a later reformulation of the theory.

*If* Coleman's observations about the 1+1-D case also apply here, the electron in such a model may be described as a tightly bound state of highly massive bosons. The work of Vachaspati [14,15] suggests that it may be possible to extend such a picture to the entire standard model of physics, resulting in a model which is well-defined even without renormalization, without any need to postulate additional unobserved dimensions of space (as in superstring theory).

If all of this can be accomplished, then our recent extensions (quant-ph 0309087) of classical-quantum statistical equivalence [16-19] suggest that there may exist a similar equivalence in the limit between the BQFT based on W(r) and the "P-mapped" version of the statistics of a family of classical field theories based on a similar path W'(r). One can unify this with general relativity, without leaving curved Minkowki space, simply by metrifying the classical field theories and analyzing the resulting statistics, in the spirit of Penrose [20].

It is far too early to say what the practical benefits of such a new theoretical foundation would be, if any. However, recent discussion of "preon" models in nuclear physics suggest that a model with r slightly larger than zero might actually fit empirical data better than the standard model. A BQFT with such an r might suggest new types of phenomena to look for, related to magnetic monopoles and such. Likewise, the use of classical-quantum correspondences [15-19] might possibly suggest modes of coherent excitations that could suspend the "mass gap" of QCD long enough to be interesting. But the key task for now is to set straight the foundations, to enable such later development.